\documentclass{elsart5p}

\pdfoutput=1
\usepackage{amsmath}
\usepackage{amsfonts}
\usepackage{amssymb}
\usepackage{graphicx}

\renewcommand{\Re}{\mathrm{Re}}
\renewcommand{\Im}{\mathrm{Im}}
 
\begin{document}

\begin{frontmatter}

\title{Effect of Cauchy noise on a network of quadratic integrate-and-fire neurons with  non-Cauchy heterogeneities}

\author{Viktoras Pyragas and Kestutis Pyragas}
\address{Center for Physical Sciences and Technology, Sauletekio al. 3, LT-10257 Vilnius, Lithuania}

\begin{abstract}
We analyze the dynamics of large networks of pulse-coupled quadratic integrate-and-fire neurons driven by Cauchy noise and non-Cauchy heterogeneous inputs. Two types of heterogeneities defined by families of $q$-Gaussian and flat distributions are considered. Both families are parametrized by an integer $n$, so that as $n$ increases, the first family tends to a normal distribution, and the second tends to a uniform distribution. For both families, exact systems of mean-field equations are derived and their bifurcation analysis is carried out.  We show that noise and heterogeneity can have qualitatively different effects on the collective dynamics of neurons. 
\end{abstract}

\begin{keyword}
Noisy neural networks; 
Mean-field reduction; 
Ott-Antonsen ansatz;
Quadratic integrate-and-fire neurons; 
Bifurcation analysis 
\end{keyword}

\end{frontmatter}

\section{\label{sec:Introduction} Introduction}

Many biological systems, such as the brain, are made up of a huge number of dynamic units. Modeling such systems at the microscopic level requires large computational resources. An alternative approach is to use simplified heuristic models on a coarse scale. In neuroscience, such models are known as neural mass models~\cite{Wilson1973,Destexhe2009}.  They successfully explain a certain class of phenomena, but cannot correctly describe synchronization processes in neural networks. Significant progress in the development of low-dimensional models capable of adequately describing the collective dynamics of large-scale neural networks has been made only recently~\cite{Schwalger2019,Bick2020,Coombes2019,Coombes2023}. The new approach is based on the consideration of synchronizing systems by methods of statistical physics~
\cite{Gupta2018}. In the thermodynamic limit of an infinite-size network, this approach allows one to derive an exact low-dimensional system of mean-field equations from the microscopic dynamics of individual neurons. Mean-field equations derived in such a way were called the next-generation neural mass models~\cite{Coombes2019,Coombes2023}. 

The general mathematical method for next-generation models was developed by Ott and Antonsen (OA)~\cite{Ott2008}. They found that the dynamics of an infinitely large population of globally coupled heterogeneous phase oscillators converges to an invariant low-dimensional manifold -- the OA manifold. A reduced system of two ordinary differential equations (ODEs) was obtained to describe the population dynamics on this manifold.  Luke {\it et al}.~\cite{Luke2013} applied this approach for a  network of theta neurons~\cite{Ermentrout1986}. Montbri\'o {\it et al}.~\cite{Montbrio2015} derived a reduced system of mean-field equations for a heterogeneous network of all-to-all pulse-coupled quadratic integrate-and-fire (QIF) neurons. QIF and theta neuron models can be transformed into each other by changing variables. They represent the canonical forms of class I neurons~\cite{izhi07}. For QIF neurons, the OA manifold transforms into the Lorentzian form~\cite{Montbrio2015}. In recent years, reduced mean-field equations have been obtained for a number of different modifications of QIF neural networks~\cite{Pazo2016,Ratas2016,Devalle2017,Ratas2017,Ratas2018,Montbrio2020,Segneri2020,Klinshov2021,PyragasV2021,PyragasV2022}. 
To simplify the reduction, it is usually assumed that the heterogeneity is Cauchy distributed, although in some publications~\cite{Klinshov2021,PyragasV2021,PyragasV2022} mean-field equations have been obtained for non-Cauchy heterogeneities.

Since noise is an important factor in neural systems~\cite{Longtin2003}, the extension of OA theory to noisy QIF neural networks is a relevant problem.
Unfortunately, there is no exact low-dimensional reduction for QIF neural populations driven by Gaussian noise, and only approximate mean-field theories are possible~\cite{Tyulkina2018,Ratas2019,GoldobinPRL2021,Goldobin2021}.
However, the situation changes when the noise is not Gaussian but Cauchy. In recent publications~\cite{Tanaka2020,Tonjes2020} it was shown that the exact low-dimensional reduction is applicable to Kuramoto oscillators driven by Cauchy white noise. Then the exact mean-field equations for the QIF neural network driven by  heterogeneous Cauchy distributed time-independent (quenched) inputs and Cauchy white noise were obtained~\cite{Clusella2022,Pietras2023}. These publications showed that Cauchy white noise and Cauchy-distributed heterogeneous inputs have the same effect on the macroscopic behavior of the network. Qualitative similarity at the macroscopic level between noise and quenched heterogeneity was also observed in the case when noise is Gaussian and heterogeneity is Cauchy distributed~\cite{Montbrio2015,Ratas2018,Strogatz1991}. A natural question arises: are the effects of noise and heterogeneity qualitatively similar in the general case, regardless of their statistics?

In this paper, we investigate the dynamics of a QIF neural network driven by Cauchy white noise and non-Cauchy heterogeneous inputs. Two types of quenched heterogeneities defined by families of $q$-Gaussian distributions~\cite{PyragasV2022,Tsallis2009} and flat distributions~\cite{Skardal2018} are considered. For both cases, we derive exact systems of mean-field equations and show that noise and heterogeneity can have qualitatively different effects on the collective dynamics of the network. Our approach, just as in Ref.~\cite{Clusella2022}, is based on the standard OA theory. A recently proposed more general approach~\cite{Pietras2023}, which can describe not only the dynamics on the OA manifold but also the transitions to this manifold, is beyond the scope of this paper.

The paper is organized as follows. Section~\ref{sec:model} describes the model and formulates the problem. In Sec.~\ref{sec:Mean-field}, we apply the OA reduction to a QIF neural network driven by Cauchy white noise and arbitrarily distributed heterogeneous inputs. We then use the general results obtained in Sec.~\ref{sec:Mean-field} to derive mean-field equations for specific heterogeneities defined by families of $q$-Gaussian distributions (Sec.~\ref{sec:q-Gaussian}) and flat distributions (Sec.~\ref{sec:Rectangular}). In Secs.~\ref{sec:q-Gaussian} and \ref{sec:Rectangular}, we also analyze the bifurcations of the corresponding mean-field equations and compare their solutions with those of microscopic models. The conclusions are presented in Sec.~\ref{sec:Conclusions}.

\section{Model and problem formulation}
\label{sec:model}

We consider a noisy heterogeneous population of $i=1,\ldots,N$  quadratic integrate-and-fire neurons \cite{izhi07} interacting via a mean-field inhibitory coupling. The evolution of the membrane potential $V_i$ of a neuron $i$ in the ensemble is described by the differential equation
\begin{eqnarray}
\tau_m\dot{V}_{i} =  {V}_{i}^{2}+\eta_i-J\tau_m S(t)+\Gamma \xi_i(t) \label{model}
\end{eqnarray}
with the auxiliary after-spike resetting rule
\begin{eqnarray}
\text{if} \; {V}_{i} \ge V_{p} \; \text{then} \; {V}_{i} \leftarrow V_{r}. \label{reset}
\end{eqnarray}
Here, $\tau_m$ is the membrane time constant, the overdot denotes the time derivative, the heterogeneous quenched parameter $\eta_i$  is a time-independent input current that specifies the behavior of each isolated neuron, $J\geq 0$ is the strength of the synaptic coupling, $S(t)$ is the synaptic activation. In addition, neurons are subject to independent noise currents $\Gamma \xi_i(t)$,  where $\xi_i(t)$ represent Cauchy white noise with zero mean and half-width at half-maximum (HWHM) equal to one. The parameter $\Gamma$ determines the intensity of the noise (the HWHM of the term $\Gamma \xi_i(t)$ is equal to $\Gamma$). The isolated neurons ($J=0$ and $\Gamma=0$) with the negative value of the quenched parameter $\eta_i<0$ are at rest, while the neurons with the positive value of the parameter $\eta_i>0$ generate instantaneous spikes. Each time a potential $V_i$ reaches the threshold value $V_p$, it is reset to the value $V_r$, and the neuron emits an instantaneous spike which contributes to the network mean firing rate 
\begin{eqnarray}
R= \frac{1}{N} \sum_{i=1}^N \sum_k \frac{1}{\tau_r}\int_{t-\tau_r}^t \delta(t'-t_i^k) d t', \label{firing_rate}
\end{eqnarray}
where $t_i^k$ is the time of the $k$th spike of the $i$th neuron, $\delta(t)$ is the Dirac delta function, and $\tau_r$ is a time window of spike events. The mean synaptic activity $S(t)$ satisfies the relaxation equation
\begin{eqnarray}
\tau_s \dot{S}=-S+R, \label{mean_rate}
\end{eqnarray}
where $\tau_s$ is the synaptic decay time constant of the inhibitory synapses.
Because of the quadratic nonlinearity, $V_i$ may reach infinity in a finite time, and this allows us to choose the threshold parameters as $V_{p}= -V_{r} = \infty$. With this choice, QIF neurons can be transformed into theta neurons~\cite{Ermentrout1986}. This is done by changing the variables 
\begin{equation}
V_i = \tan(\theta_i/2). \label{eq_transf_tet}
\end{equation}
For theta neurons, the Eqs.~\eqref{model} read:
\begin{equation}
\tau_m \dot{\theta}_{i} = 1-\cos \left(\theta_{i}\right)
+\left[1+\cos \left(\theta_{i}\right)\right]\left[\eta_{i}-J\tau_m S(t) +\Gamma \xi_i(t) \right]. \label{theta_j}
\end{equation}
In Eqs.~\eqref{model} noise is additive, while here it is multiplicative (in the Stratonovich interpretation). Theta neuron representation avoids discontinuities associated with instantaneous reset whenever the membrane potential $V$ of QIF neuron crosses the threshold at infinity and emits a spike. The dynamics of the theta neuron at this point remains bounded and smooth. The spike occurs when the phase variable $\theta$ passes through $\pi$.

The microscopic model described above was proposed in Ref.~\cite{Devalle2017} as a model of interneuronal gamma (ING) oscillations  \cite{Wang1996,Wang2010}. In the absence of noise, the mean-field equations for this model were derived for the Cauchy heterogeneity in \cite{Devalle2017} and extended to the q-Gaussian heterogeneity in~\cite{PyragasV2022}. Recently, exact mean-field equations have also been obtained in the presence of noise, when both the noise and the quenched heterogeneity are Cauchy distributed~\cite{Clusella2022,Pietras2023}. It turned out that noise and quenched heterogeneity in such a model have the same effect on population dynamics.

Here we analyze this model for the case of Cauchy noise and non-Cauchy quenched heterogeneity. We consider two families of distributions for the parameter $\eta$. Both of them are bell-shaped and symmetrical with respect to the maximum. We denote the mean value and HWHM of the distribution of the parameter $\eta$  as $\bar{\eta}$ and $\Delta$, respectively. To simplify the description, we replace the quenched parameter $\eta$ by $\zeta$, 
\begin{equation}
\eta=\bar{\eta}+\Delta\zeta, \label{zeta}
\end{equation}
so that the distribution $f(\zeta)$ of the modified quenched parameter $\zeta$ has zero mean and HWHM equal to one. 

\begin{figure}
\centering\includegraphics{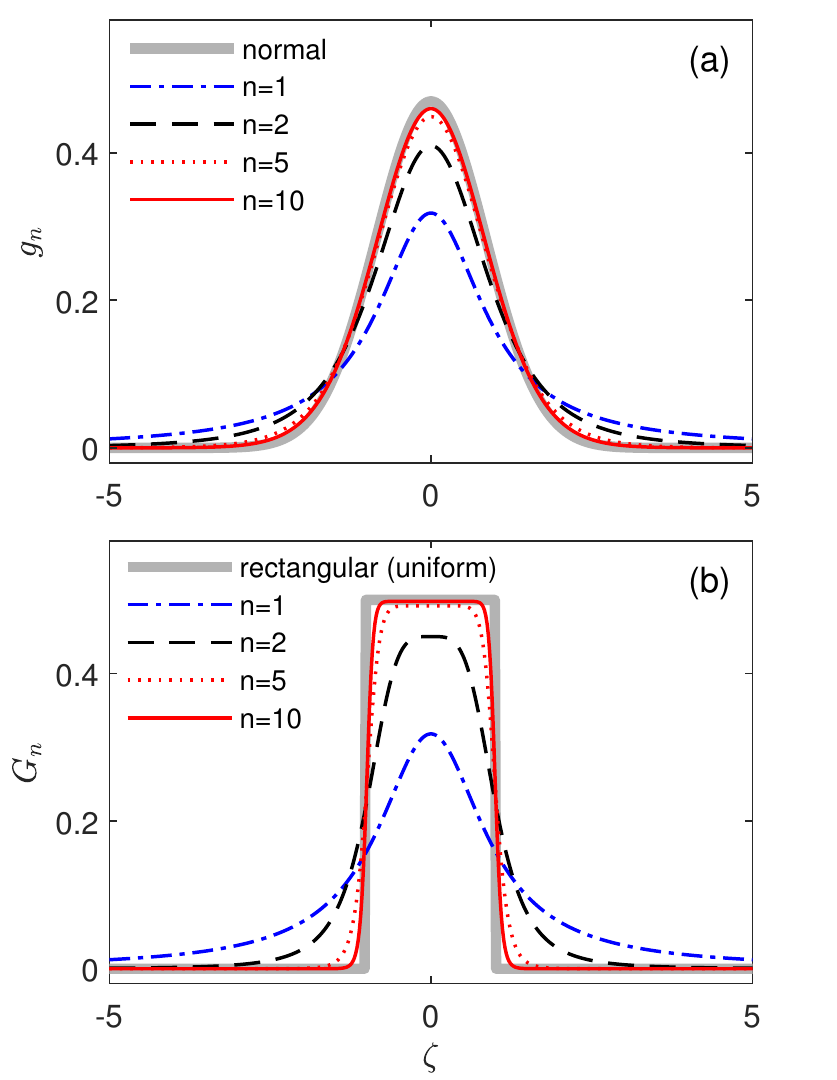}
\caption{\label{fig:distrib} Distributions of the heterogeneous input currents. (a) $q$-Gaussian distribution \eqref{n-Gaussian} for different values of MTI $n$. For $n=1$ it coincides with the Cauchy distribution~\eqref{Cauchy} and for $n\to \infty$ it tends to the normal distribution~\eqref{Gauss}. (b) Flat distribution~\eqref{n-Rect}.  For $n=1$, it coincides with the Cauchy distribution, and for $n\to \infty$, it tends to the rectangular (uniform) distribution~\eqref{Rect}. }
\end{figure}

The first family of distributions that we consider in this paper is the restricted class~\cite{PyragasV2022} of $q$-Gaussian density functions introduced by Tsallis in non-extensive statistical mechanics~\cite{Tsallis2009}: 
\begin{equation}
f(\zeta)=g_n(\zeta) = \frac{c_n}{(1+\beta_n \zeta^2)^n}, \label{n-Gaussian}
\end{equation}
where $n$ is a positive integer called the modified Tsallis  index (MTI)~\cite{PyragasV2022}, 
\begin{equation}
c_n =\sqrt{\frac{\beta_n}{\pi}} \frac{\Gamma(n)}{\Gamma(n-1/2)}, \label{c_n}
\end{equation}
is the normalization constant and $\Gamma(\cdot)$ is the Gamma function. The factor 
\begin{equation}
\beta_n =2^{1/n}-1  \label{beta_n}
\end{equation}
ensures that HWHM of the distribution \eqref{n-Gaussian} is equal to one for all $n=1,2,\ldots,\infty$, i.e., $g_n(1)=g_n(0)/2$. For $n=1$, this distribution coincides with the Cauchy distribution 
\begin{equation}
g_1(\zeta)= \frac{1}{\pi}\frac{1}{1+\zeta^2}, \label{Cauchy}
\end{equation}
and for $n \to \infty$ it tends to the normal Gaussian distribution:
\begin{equation}
g_\infty(\zeta)=\sqrt{\frac{\ln(2)}{\pi}} \exp\left[-\zeta^2\ln(2)\right]. \label{Gauss}
\end{equation}
The evolution of the distribution with increasing $n$ is shown in Fig.~\ref{fig:distrib}(a). For $n \sim 10$, the $q$-Gaussian distribution is close to the normal distribution.

The second family of distributions considered in this paper is a family of flat distributions defined by a rational function~\cite{Skardal2018}:
\begin{equation}
f(\zeta)=G_n(\zeta) = \frac{C_n}{1+\zeta^{2n}}, \label{n-Rect}
\end{equation}
where $n$ is again a positive integer, and the normalization constant is  
\begin{equation}
C_n =\frac{n}{\pi} \sin\left(\frac{\pi}{2n}\right). \label{C_n}
\end{equation}
The distribution \eqref{n-Rect} was used in~\cite{Skardal2018} to describe the heterogeneity of natural frequencies of phase oscillators in the Kuramoto model. 
As in the previous case, the HWHM of this distribution is equal to one for all $n$ and, for $n=1$, it coincides with the Cauchy distribution, $G_1(\zeta)=g_1(\zeta)$. But for $n\to \infty$, this distribution tends to the rectangular (uniform) distribution:
\begin{equation}
G_\infty(\zeta) =
\begin{cases}
1/2 &\text{ for } |\zeta| <1\\
0 &\text{ for } |\zeta|>1
\end{cases}
\label{Rect}
\end{equation}
The evolution of the distribution~\eqref{n-Rect} with increasing $n$ is shown in Fig.~\ref{fig:distrib}(b). For $n \sim 10$, it is close to the rectangular distribution.

\section{Ott-Antonsen reduction for Cauchy noise and arbitrary heterogeneity }
\label{sec:Mean-field}
 
The Ott-Antonsen reduction method~\cite{Ott2008} has been originally discovered for the noiseless Kuramoto model and then applied to noiseless QIF and theta neuron systems. Obtaining reduced models for large populations of neurons in the presence of noise is an actual problem in theoretical neuroscience. In the case of Gaussian noise, the exact Ott-Antonsen reduction method does not work, but for small noises, one can still derive approximate low-dimensional models~\cite{Tyulkina2018,Ratas2019,GoldobinPRL2021,Goldobin2021}. 
A surprising recent discovery was that exact low-dimensional reduction is applicable in the case of Cauchy noise. Such a reduction was demonstrated for the Kuramoto oscillators~\cite{Tanaka2020,Tonjes2020} and  QIF (or theta) neurons~\cite{Clusella2022,Pietras2023}  when both noise and heterogeneity are Cauchy distributed. Here we briefly reproduce these results without requiring that the quenched heterogeneity be Cauchy distributed. Following Ref.~\cite{Clusella2022}, we use standard OA theory and generalize the results of this paper to the case of arbitrarily distributed heterogeneity.

In the thermodynamic limit $N\to \infty$, the  macroscopic
state of theta neurons can be characterized  by the probability
density function of neurons having  the phase $\theta$ at time $t$,
\begin{equation}
 \rho(\theta,t)= \int_{-\infty}^\infty d\zeta f(\zeta) P(\theta,t;\zeta),
\label{Prob}
\end{equation}
where  $P(\theta,t;\zeta)$ are  conditional densities for specific values
of $\zeta$. In the case of Cauchy noise, the evolution of the function $P(\theta,t;\zeta)$ satisfies the fractional Focker-Plank equation: 
%
\begin{eqnarray}
    \tau_m \frac{\partial}{\partial t} P= &-&\frac{\partial}{\partial \theta}\left\{ \left[ 1-\cos\theta+(1+\cos\theta)(\eta-J\tau_m S) \right]P \right\}\nonumber\\
    &+&\Gamma\frac{\partial}{\partial |\theta|}\left[ (1+\cos\theta) P \right].
    \label{FP_eq}
\end{eqnarray}
%
Here the operator $\frac{\partial}{\partial |\theta|}$ denotes the Riesz fractional derivative that in the Fourier space acts as $\frac{\partial}{\partial |\theta|}e^{ik\theta}=-|k|e^{ik\theta}$. We expand the probability density into Fourier series 
\begin{equation}
 P(\theta,t;\zeta)= \frac{1}{2\pi}\left(1+\sum_{k=1}^{\infty} p_k(t,\zeta) e^{ik\theta}+\text{c.c.}\right)
\label{Four_exp}
\end{equation}
with Fourier coefficients
\begin{equation}
p_k(t,\zeta)=\int_0^{2\pi} d\theta P(\theta,t;\zeta)e^{-ik\theta}.
\label{Four_coef}
\end{equation}
Substituting the Eq.~\eqref{Four_exp} into Eq.~\eqref{FP_eq}, we get
an infinite set of differential equations for the Fourier coefficients, $k\geq 1$,
\begin{eqnarray}
   \tau_m \dot{p}_k=&-& k\left[i(\eta-J\tau_m S+1)+\Gamma  \right]p_k\nonumber\\
    &-& k\left[i(\eta-J\tau_m S-1)+\Gamma  \right](p_{k+1}+p_{k-1})/2.
    \label{pk_eq}
\end{eqnarray}
The coefficient $p_0(t,\zeta)\equiv 1$ due to the normalization of the density function. With the Ott-Antonsen anzatz~\cite{Ott2008} 
\begin{equation}
p_k(t,\zeta)=[\alpha(t,\zeta)]^k,
\label{OA_anzatz}
\end{equation}
the infinite set of differential Eqs.~\eqref{pk_eq} reduces to a single equation for the complex parameter $\alpha(t,\zeta)$:
\begin{eqnarray}
   \tau_m \dot{\alpha} = &-&\left[i(\bar{\eta}+\Delta\zeta-J\tau_m S+1)+\Gamma  \right]\alpha \nonumber\\
    &-& \left[i(\bar{\eta}+\Delta\zeta-J\tau_m S-1)+\Gamma  \right](1+\alpha^2)/2.
    \label{alp_eq}
\end{eqnarray}
%
Substituting the Eq.~\eqref{OA_anzatz} into Eq.~\eqref{Four_exp}, we find the expression of the density function on the OA manifold
\begin{equation}
 P(\theta,t;\zeta)= \frac{1}{2\pi}\Re \left[\frac{1+\alpha(t,\zeta)e^{i\theta}}{1-\alpha(t,\zeta)e^{i\theta}}\right].
\label{Prob_OA}
\end{equation}
In the voltage (QIF) representation, $V=\tan(\theta/2)$, the density function on the OA manifold takes the form of the Lorentzian function~\cite{Montbrio2015}  
\begin{eqnarray}
F(V,t;\zeta)&=&P\left(2\arctan(V),t;\zeta \right)\frac{2}{1+V^2}\nonumber
\\&=&\frac{1}{\pi}\frac{x(t,\zeta)}{\left[V-y(t,\zeta)\right]^2+\left[x(t,\zeta)\right]^2},
\label{Prob_Lor}
\end{eqnarray}
where $x(t,\zeta)$ and $y(t,\zeta)$ are the real and imaginary parts of the complex parameter $w(t,\zeta)=x(t,\zeta)+i y(t,\zeta)$ that is related to the parameter $\alpha(t,\zeta)$ as
\begin{equation}
 w(t,\zeta)= \frac{1-\alpha(t,\zeta)}{1+\alpha(t,\zeta)}.
\label{w_alp}
\end{equation}
Substituting the inverse transformation $\alpha(t,\zeta)=[1-w(t,\zeta)]/[1+w(t,\zeta)]$ into Eq.~\eqref{alp_eq} leads to a simpler dynamic equation for the parameter $w(t,\zeta)$:
\begin{equation}
\tau_m \dot{w}(t,\zeta)= i\left[\bar{\eta}-i\Gamma+\Delta\zeta-J\tau_m S(t) -w^2(t,\zeta) \right].
    \label{w_eq}
\end{equation}

In order to close the system of equations obtained in the thermodynamic limit, we need to express the mean firing rate~\eqref{firing_rate} in terms of the parameter $w(t,\zeta)$. In the limit $N\to \infty$ and $\tau_r \to 0$, the firing rate for fixed $\zeta$ can be estimated as probability flux of theta neurons at the point $\theta=\pi$: $(2/\tau_m)\cdot P(t,\pi;\zeta)=\Re[w(t,\zeta)]/(\pi\tau_m)$. Averaging this flux over $\zeta$, we obtain the mean firing rate Eq.~\eqref{firing_rate}  in the form   
\begin{equation}\label{R_t}
    R(t)=\frac{1}{\pi \tau_m}\Re[W(t)],
\end{equation}
where
\begin{equation}\label{W_t}
    W(t)=\int^{+\infty}_{-\infty} d\zeta f(\zeta) w(t,\zeta).
\end{equation}
Similarly, averaging the variable $y(t,\zeta)=\Im [w(t,\zeta)]$ over $\zeta$, we obtain the mean membrane potential $\bar{V}(t)=\Im [W(t)]$.

Equations~\eqref{w_eq}--\eqref{W_t}  and \eqref{mean_rate} form a closed system of integro-differential equations describing the dynamics of a neural population in the thermodynamic limit. Further simplification of this system is possible if the distribution $f(\zeta)$ is a rational function. In this case, the integral~\eqref{W_t} can be estimated analytically using residue theory, and the system of integro-differential equations can be reduced to a finite set of ODEs.
Next, we present such a reduction for both types of the above-mentioned heterogeneities, determined by the $q$-Gaussian distribution~\eqref{n-Gaussian} and the flat distribution~\eqref{n-Rect}. 

It is noteworthy that in the thermodynamic limit equations with Cauchy noise and arbitrary heterogeneity can be obtained from equations without noise by formally replacing the parameter $\bar{\eta}$ by $\bar{\eta}-i\Gamma$ [see Eq.~\eqref{w_eq}]. We will use this remarkable property in Sec.~\ref{sec:q-Gaussian} when considering a neural population with $q$-Gaussian heterogeneity. In the absence of noise, the reduced mean-field equations for this problem were derived in our previous publication~\cite{PyragasV2022},  and we simply use the above parameter change to supplement these equations with Cauchy noise. Heterogeneity in the form of flat distributions~\eqref{n-Rect}  has not yet been discussed in the literature for populations of QIF neurons, and in Sec.~\ref{sec:Rectangular} we will consider this case in more detail. 

\section{\label{sec:q-Gaussian} Effect of Cauchy noise on a neural population with $q$-Gaussian heterogeneity}

As mentioned above, the mean-field equations for a population of QIF neurons with $q$-Gaussian heterogeneity and no noise were derived in~\cite{PyragasV2022}. The main point of this derivation was the evaluation of the integral~\eqref{W_t} for the function~\eqref{n-Gaussian}, $f(\zeta)=g_n(\zeta) = c_n/(1+\beta_n \zeta^2)^ n$, which has two $n$-order poles $\zeta=\pm i \beta_n^{-1/2}$ on the complex plane $\zeta$. Using  residue theory for higher-order poles, this integral was expressed as a linear combination of $n$ time-dependent order parameters $W_1(t),\ldots, W_n(t)$,
\begin{equation}\label{W_t_3}
W(t)=\sum_{k=1}^{n}b_{k}W_{k}(t),
\end{equation}
whose dynamics satisfy a system of $n$ ODEs presented below, and the
coefficients $b_k$ are given by the recurrence relation
\begin{subequations}\label{b_k_4}
\begin{eqnarray}
b_{1}&=&1,\\
b_{k}&=&\frac{n-k+1}{n-k/2} b_{k-1}, \quad k=2,\ldots,n.
\end{eqnarray}
\end{subequations}
Differential equations for the order parameters $W_k(t)$ were derived in ~\cite{PyragasV2022} from the Eq.~\eqref{w_eq} for $\Gamma=0$. To account for Cauchy noise, here we rewrite these equations with the substitution $\bar{\eta} \leftarrow \bar{\eta} -i\Gamma$. As a result, we obtain the following closed system of $N+1$ ODEs
\begin{subequations}\label{w_k_ode}
\begin{eqnarray}
\tau_{m}\dot{W}_{1}&=&i[\bar{\eta}-i\Gamma-i\Delta\beta_n^{-1/2}-J\tau_m S-W_{1}^{2}],\label{w_k_odea}\\
\tau_{m}\dot{W}_{2}&=& -\Delta \beta_n^{-1/2}-i2W_1W_2,\label{w_k_odeb}\\
\tau_{m}\dot{W}_{k}&=&-i\sum_{l=1}^{k}W_{k-l+1}W_{l}, \quad k=3,\ldots,n, \label{w_k_odec}\\
\tau_s \dot{S}&=&-S+R \label{w_k_oded}
\end{eqnarray}
\end{subequations}
that describes the dynamics of QIF neurons with $q$-Gaussian heterogeneity and Cauchy noise. Here the firing rate $R(t)$ is determined by the Eqs.~\eqref{R_t} and \eqref{W_t_3}.

For $n=1$, the $q$-Gaussian distribution coincides with the Cauchy distribution,  and in this case, Eqs.~\eqref{w_k_ode} are the same as the equations discussed in ~\cite{Clusella2022}. The system dynamics is determined by only two ODEs~\eqref{w_k_odea} and \eqref{w_k_oded} with $\beta_1=1$ and $W(t)=W_1(t)$. We see that the quantities $\Delta$ and $\Gamma$, which determine the HWHM of the quenched heterogeneity and the noise intensity, respectively, appear in these equations as the sum $\Delta+\Gamma$. Thus, as stated in~\cite{Clusella2022}, both Cauchy noise and Cauchy quenched heterogeneity has the same effect on population dynamics. However, for $n>1$, the parameters $\Delta$ and $\Gamma$ appear differently in Eqs.~\eqref{w_k_ode}, and in the general case, one can expect different effects from noise and quenched heterogeneity.

Next, we analyze the influence of  $\Delta$ and $\Gamma$ parameters on population dynamics depending on the MTI $n$. Before proceeding to this analysis, let us reduce the number of parameters in the Eqs.~\eqref{w_k_ode} by rewriting them in a dimensionless form, similar to what was done in~\cite{PyragasV2022}. Assuming $\bar{\eta} > 0$, we introduce the dimensionless time
\begin{equation}\label{vartheta}
\vartheta= t\sqrt{\bar{\eta}}/\tau_m  
\end{equation}
and change the variables as
\begin{equation}\label{Dim_les_var}
w_{k}=W_{k}/\sqrt{\bar{\eta}}, \quad s=S\tau_{m}/\sqrt{\bar{\eta}},\quad r=R\tau_{m}/\sqrt{\bar{\eta}}.
\end{equation}
Then the system~\eqref{w_k_ode} takes the form
\begin{subequations}\label{m_f_eq}
\begin{eqnarray}
w'_{1}&=&i[1-i\gamma-i\delta \beta_n^{-1/2}-js-w_{1}^{2}],\label{m_f_eqa}\\
w'_{2}&=& -\delta \beta_n^{-1/2}-i2w_1w_2,\label{m_f_eqb}\\
w'_{k}&=&-i\sum_{l=1}^{k}w_{k-l+1}w_{l}, \quad k=3,\ldots,n, \label{m_f_eqc}\\
\tau s'&=&-s+r, \quad r=\frac{1}{\pi}\Re\sum_{l=1}^{n}b_{l}w_{l}, \label{m_f_eqd}
\end{eqnarray}
\end{subequations}
where the prime denotes the derivative with respect to the dimensionless time $\vartheta$. The system~\eqref{m_f_eq} contains only four independent parameters:
\begin{equation}
j=J/\sqrt{\bar{\eta}},\quad \tau=\sqrt{\bar{\eta}}\tau_{s}/\tau_{m},\quad \delta=\Delta/\bar{\eta}, \quad \gamma=\Gamma/\bar{\eta}. \label{param}
\end{equation}
The parameter $j$ is the new normalized coupling strength, $\tau$ is proportional to the ratio of the synaptic time constant $\tau_s$ to the membrane time constant $\tau_m$, $\delta$ is the ratio of the HWHM $\Delta$ to the center $\bar{\eta}$ of the quenched parameter $\eta$, and $\gamma$ is the ratio of the noise intensity $\Gamma$ to $\bar{\eta}$. 

\begin{figure}
\centering\includegraphics{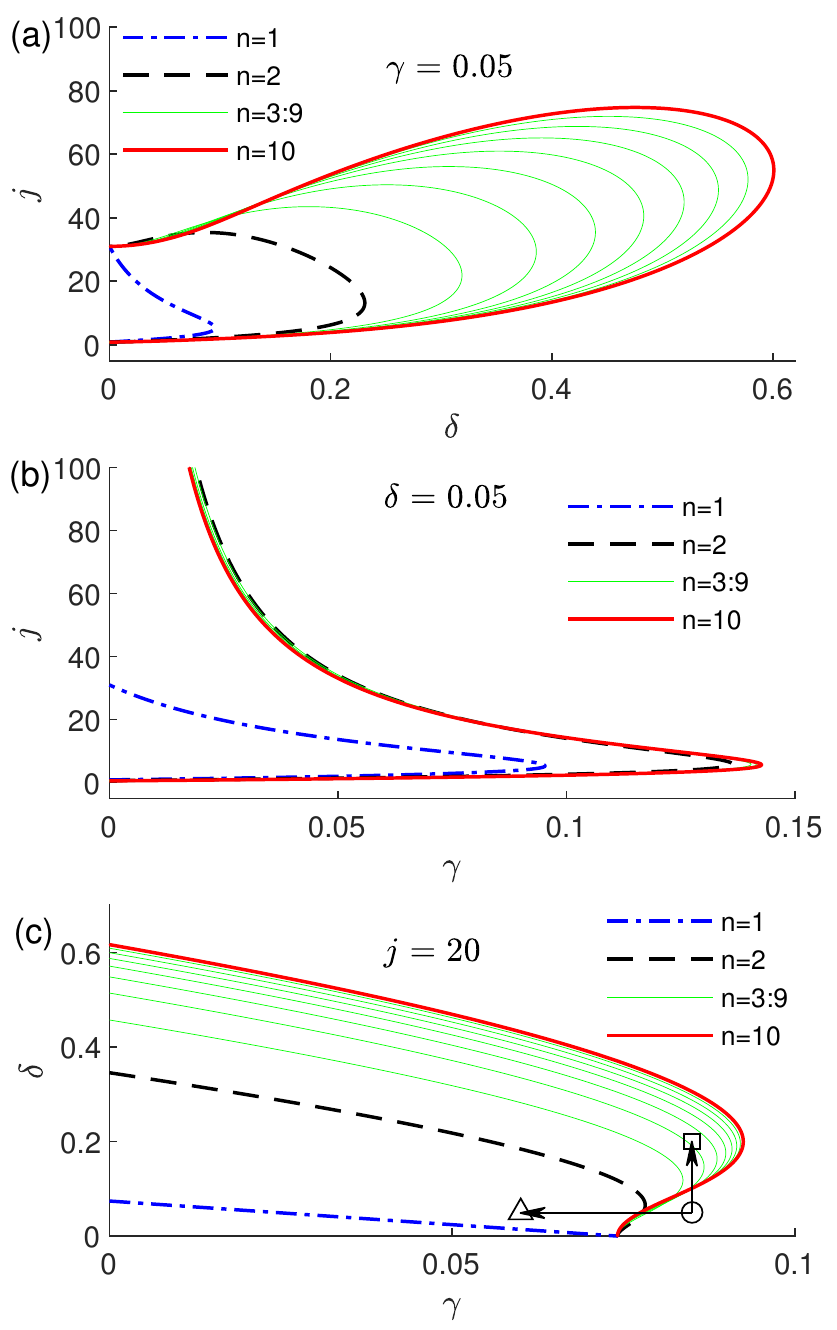}
\caption{\label{fig:bif_gauss} Two-parameter bifurcation diagrams of a population of QIF neurons with $q$-Gaussian heterogeneity and Cauchy noise in the planes of parameters (a) $(\delta, j)$, (b) $(\gamma, j)$ and (c) $(\gamma,\delta)$. Lines of different styles indicate Hopf bifurcation curves with different values of the MTI $n$. The parameter $\tau=1$ is the same for all diagrams. The corresponding value of the remaining independent parameter is displayed on each diagram. In diagram (c), circle, triangle, and square symbols indicate fixed values of parameters $(\gamma,\delta)$, equal to $(0.085,0.05)$, $(0.06,0.05)$ and $(0.085,0.2)$, respectively.
}
\end{figure}

Figures~\ref{fig:bif_gauss}(a), \ref{fig:bif_gauss}(b) and \ref{fig:bif_gauss}(c) show bifurcation diagrams on parameter planes $(\delta,j) $, $(\gamma,j)$ and $(\gamma,\delta)$, respectively, calculated from the mean-field Eqs.~\eqref{m_f_eq} using the MATCONT package~\cite{matcont}. The parameter $\tau=1$ is the same for all diagrams. Lines of different styles indicate Hopf bifurcation curves with different values of $n$. In the respective parameter planes, they limit the regions with stable limit cycles. Outside these regions, there are stable incoherent equilibrium states. In Fig.~\ref{fig:bif_gauss}(a) [parameter plane $(\delta,j)$], we take $\gamma=0.05$, and in Fig.~\ref{fig:bif_gauss}(b) [parameter plane $(\gamma,j)$], we take the same value for $\delta=0.05$. As expected, under the Cauchy heterogeneity ($n=1$), the Hopf bifurcation curves in these figures coincide. The equivalent effect of changing the parameters $\gamma$ or $\delta$ for Cauchy heterogeneity is most evident in Fig.~\ref{fig:bif_gauss}(c): in the parameter plane $(\gamma,\delta)$, the Hopf bifurcation curve for $n=1$ is a line $\gamma+\delta=\text{const.}$ 

For $n>1$, parameters $\gamma$ and $\delta$ affect the dynamics of the system in different ways. The Hopf bifurcations curves in Figs.~\ref{fig:bif_gauss}(a) and~\ref{fig:bif_gauss}(b) differ significantly for $n>1$. Limit cycle oscillations are less sensitive to an increase in quenched heterogeneity than to an increase in noise intensity. For example, for $n=10$ and a fixed $\gamma=0.05$, the limit cycle oscillations persist as the heterogeneity parameter increases to the value of $\delta \approx 0.6$, while for $n=10$ and a fixed $ \delta =0.05$, oscillations of the limit cycle are preserved only at noise intensity $\gamma \leq 0.14$. 

Even more surprisingly, changes in noise and heterogeneity in opposite directions can lead to the same bifurcation at the macroscopic level. In particular, collective limit cycle oscillations can be induced by either a reduction in noise or an increase in quenched heterogeneity. This effect is visible in Fig.~\ref{fig:bif_gauss}(c). Let's say we are dealing with MTI $n=10$. Then the Hopf bifurcation curve, which limits the region of limit cycle oscillations, is a thick red curve. Assume that initially, the parameters are outside this region, at the point $(\gamma,\delta)=(0.085,0.05)$ shown by the circle. Then one can enter the region of limit cycle oscillations, either by reducing the noise intensity, for example, by going to the point $(\gamma,\delta)=(0.06,0.05)$, shown by a triangle, or by increasing the heterogeneity, for example, by going to the point $(\gamma,\delta)=(0.085,0.2)$ shown as a square. 

\begin{figure}
\centering\includegraphics{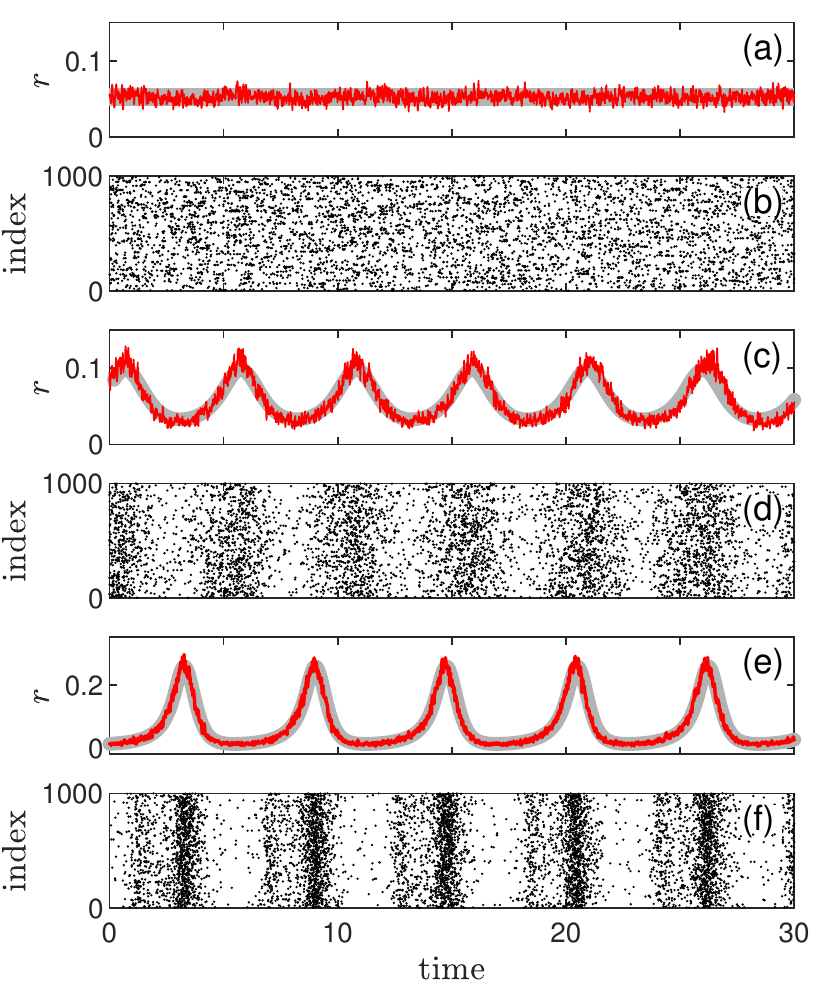}
\caption{\label{fig:dynamics_gauss} Dynamics of a population of normally distributed $5 \times 10^4$ QIF neurons driven by Cauchy noise and its comparison with the solutions of mean-field equations. (a), (c) and (e) Time traces of firing rates for parameters $(\gamma,\delta)$, equal to $(0.085,0.05)$, $(0.06,0.05)$ and $(0.085,0.2)$, which are indicated in Fig.~\ref{fig:bif_gauss}(c) by circle, triangle and square symbols, respectively. Thin red curves represent the solutions of the microscopic model Eqs.~\eqref{theta_j} with normally distributed heterogeneity, and thick gray curves are the solutions of mean-field Eqs.~\eqref{m_f_eq} for $n=10$. (b), (d) and (f) Raster plots of 1000 randomly selected neurons that complement panels (a), (c), and (e), respectively. Here dots show the spike moments for each neuron, where the vertical axis indicates neuron numbers. Time is presented in dimensionless units defined by Eq.~\eqref{vartheta}. Other parameters, $j=20$ and $\tau=1$, are the same as in Fig.~\ref{fig:bif_gauss}(c).   
}
\end{figure}

The thick gray curves in Figs.~\ref{fig:dynamics_gauss}(a), \ref{fig:dynamics_gauss}(c) and \ref{fig:dynamics_gauss}(f) show the dynamics of the firing rate, calculated from the mean-field Eqs.~\eqref{m_f_eq} for the values of the parameters indicated in Fig.~\ref{fig:bif_gauss}(c) by a circle, a triangle and a square, respectively. The thin red curves in these figures show the firing rates obtained from the microscopic model Eqs.~\eqref{theta_j} using $N=5\times10^4$ neurons with a normally distributed excitability parameter. Figures 
\ref{fig:dynamics_gauss}(b), \ref{fig:dynamics_gauss}(d) and \ref{fig:dynamics_gauss}(e) show the corresponding raster plots of 1000 randomly selected neurons. The Eqs.~\eqref{theta_j} were integrated by the Euler method with a time step in dimensionless units \eqref{vartheta} equal to $10^{-3}$. The time window $\tau_r$ of spike events in the Eq.~\eqref{firing_rate} in dimensionless units is taken equal to $10^{-2}$. Cauchy white noise was simulated as described in Ref.~\cite{Tanaka2020}. We see that the mean-field Eqs.~\eqref{m_f_eq} with the MTI $n=10$ describe well the dynamics of the mean firing rate of $N=5\times10^4$ QIF noisy neurons with normally distributed heterogeneity.   

\section{\label{sec:Rectangular}  Effects of Cauchy noise on a neural population with a flat heterogeneity distribution}

To derive mean-field equations for a flat distribution of quenched heterogeneity we need to analytically evaluate the integral~\eqref{W_t} with the function~\eqref{n-Rect}, $f(\zeta)=G_n(\zeta) = C_n/(1+\zeta^{2n})$. This can be done by using the residue theory. Namely, function $w(t,\zeta)$ is analytically continued into a complex-valued $\zeta$, and the integration contour is closed in the lower half-plane. The function $G_n(\zeta)$ has $n$ simple poles  
\begin{equation}\label{zeta_k}
\zeta_k=\exp\left[-\frac{i\pi}{2n}(2k-1)\right], \quad k=1,\ldots,n  
\end{equation}
in the lower half-plane and $n$ complex conjugate poles $\zeta_k^*$ in the higher half-plane. Since the value of the integral~\eqref{W_t} is determined by the poles~\eqref{zeta_k} in the lower-half plane, we obtain
\begin{equation}\label{W_t_Rect}
W(t)=i\sin\left(\frac{\pi}{2n}\right)\sum_{k=1}^{n}\zeta_{k}W_{k}(t),
\end{equation}
where $W_k(t)=w(t,\zeta_k)$, $k=1,\ldots,n$, are $n$ time dependent order parameters. Differential equations for these parameters are easily derived from the Eq.~\eqref{w_eq}:
\begin{equation}
\tau_m \dot{W_k}= i\left[\bar{\eta}-i\Gamma+\Delta\zeta_k-J\tau_m S(t) -W_k^2 \right]
    \label{W_k_Recr}
\end{equation}
for $k=1,\ldots,n$. These equations together with the Eqs.~\eqref{mean_rate}, \eqref{R_t} and \eqref{W_t_Rect} constitute a closed system of $n+1$ ODEs that accurately describe averaged dynamics of noisy QIF neurons with a flat heterogeneity distribution. Using the dimensionless time \eqref{vartheta} and change of variables as in the Eq.~\eqref{Dim_les_var}, this system is reduced to the following dimensionless form:
\begin{subequations}\label{m_f_eq_Rect}
\begin{eqnarray}
w'_{k}&=& i\left[1-i\gamma+\delta\zeta_k-js -w_k^2 \right], \ k=1,\ldots,n,\label{m_f_eq_Recta}\\
\tau s'&=&-s+r, \quad r=-\frac{1}{\pi}\sin\left(\frac{\pi}{2n}\right)\Im\sum_{l=1}^{n}\zeta_{l} w_{l}. \label{m_f_eq_Rectab}
\end{eqnarray}
\end{subequations}
Here the prime denotes the derivative with respect to the dimensionless time $\vartheta$, and the complex-valued coefficients $\zeta_k$ are defined in Eq.~\eqref{zeta_k}. As in the previous case, this system contains only four independent parameters $j$, $\tau$, $\delta$ and $\gamma$ defined by the Eq.~\eqref{param}. 

\begin{figure}
\centering\includegraphics{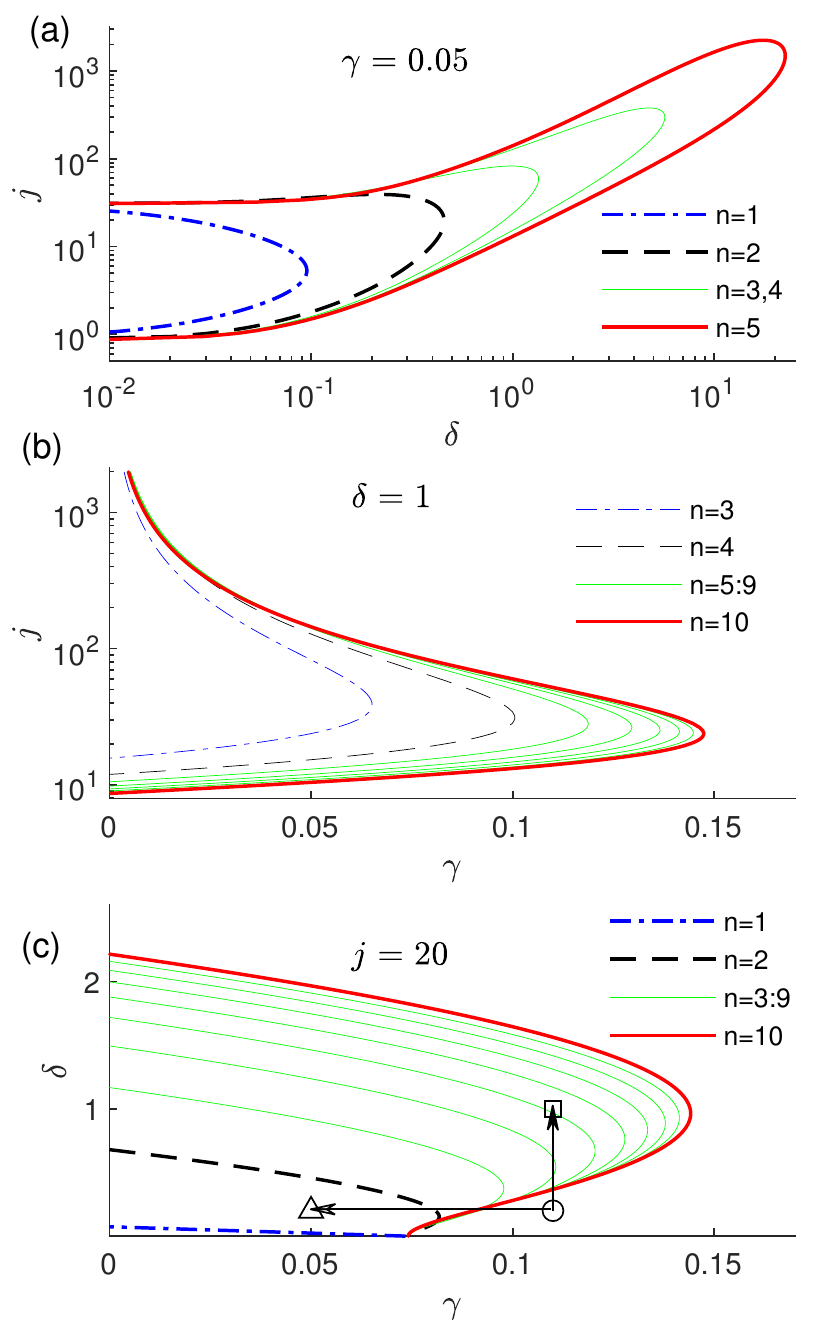}
\caption{\label{fig:bif_rect} Two-parameter bifurcation diagrams of a population of QIF neurons with a flat heterogeneity distribution \eqref{n-Rect} and Cauchy noise in the planes of parameters (a) $(\delta, j)$, (b) $(\gamma, j)$ and (c) $(\gamma,\delta)$. Lines of different styles indicate Hopf bifurcation curves with different values of $n$. The parameter $\tau=1$ is the same for all diagrams. In diagram (c), circle, triangle, and square symbols indicate fixed values of parameters $(\gamma,\delta)$, equal to $(0.11,0.2)$, $(0.05,0.2)$ and $(0.11,1.0)$, respectively.
}
\end{figure}

Figure~\ref{fig:bif_rect} shows two-parameter bifurcation diagrams for flat heterogeneity distribution similar to those shown in Fig.~\ref{fig:bif_gauss}. Limit cycle oscillations now take place in a wider range of parameters  $\delta$ and $j$. For this reason, here the bifurcation diagram ($\delta,j$) is presented on a double logarithmic scale. However, the critical amplitude of the noise that destroys the limit cycle is comparable for $q$-Gaussian and flat distributions. In both cases, coherent oscillations are impossible when the noise amplitude  $\gamma > 0.15$ [cf. Figs.~\ref{fig:bif_rect}(b) and \ref{fig:bif_gauss}(b)]. Bifurcation diagrams in the parameter plane $(\gamma,\delta)$ for both types of distributions are also qualitatively similar [cf. Figs.~\ref{fig:bif_rect}(c) and \ref{fig:bif_gauss}(c)].

\begin{figure}
\centering\includegraphics{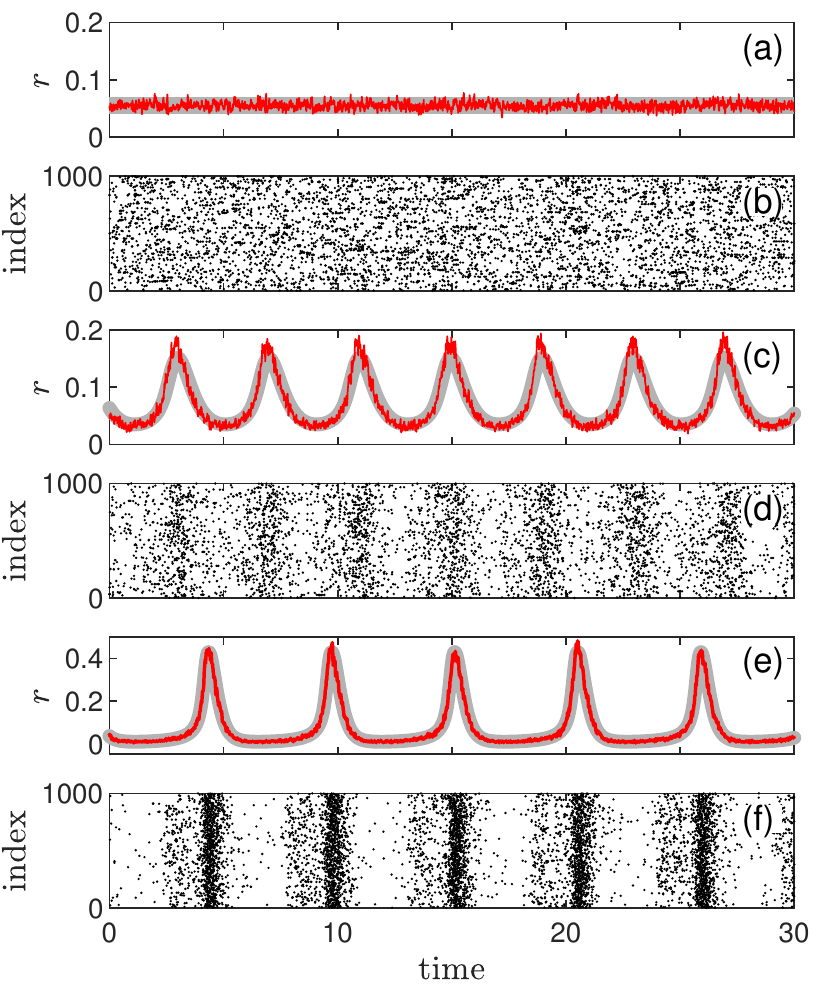}
\caption{\label{fig:dynamics_rect} Dynamics of a population of uniformly distributed $5 \times 10^4$ QIF neurons driven by Cauchy noise and its comparison with the solutions of mean-field equations. (a), (c) and (e) Time traces of firing rates for parameters $(\gamma,\delta)$, equal to $(0.11,0.2)$, $(0.05,0.2)$ and $(0.11,1.0)$, which are indicated in Fig.~\ref{fig:bif_rect}(c) by circle, triangle and square symbols, respectively. Thin red curves represent the solutions of the microscopic model Eqs.~\eqref{theta_j} with uniformly distributed heterogeneous inputs and thick gray curves are the solutions of mean-field Eqs.~\eqref{m_f_eq_Rect} for $n=10$. (b), (d) and (f) Raster plots of 1000 randomly selected neurons that complement panels (a), (c), and (e), respectively. Time is presented in dimensionless units defined by Eq.~\eqref{vartheta}. Other parameters, $j=20$ and $\tau=1$, are the same as in Fig.~\ref{fig:bif_rect}(c).   
}
\end{figure}

Figure~\ref{fig:bif_rect}(c) shows that in the case of a flat distribution, changes in noise and heterogeneity in opposite directions can also lead to qualitatively the same effect. Consider $n=10$ when the distribution \eqref{n-Rect} is close to uniform. The Hopf bifurcation curve, which limits the range of limit cycle oscillations in the parameter plane $(\gamma,\delta)$, is shown by a thick red line. For parameters outside this region, at the point $(\gamma,\delta)=(0.11,0.2)$ shown by the circle, the population is in a stable incoherent state. The dynamics of the system in this state, calculated using the Eqs.~\eqref{theta_j} of the microscopic model with the uniform distribution and the mean-field Eqs.~\eqref{m_f_eq_Rect} with $n=10$, is shown in Figs.~\ref{fig:dynamics_rect}(a) and \ref{fig:dynamics_rect}(b). From this state, one can enter the region of limit cycle oscillations, either by reducing the noise intensity, for example, by passing to the point $(\gamma,\delta)=(0.05,0.2)$, shown by a triangle, or by increasing the heterogeneity, for example, by passing to the point $(\gamma,\delta)=(0.11,1.0)$ shown as a square. The dynamics of the system in the state indicated by the square, calculated from the microscopic model and mean-field equations, is shown in Figs.~\ref{fig:dynamics_rect}(c) and \ref{fig:dynamics_rect}(d). The dynamics corresponding to the state indicated by the triangle are shown in Figs.~\ref{fig:dynamics_rect}(e) and \ref{fig:dynamics_rect}(f). As is seen in Figs.~\ref{fig:dynamics_rect}(a), \ref{fig:dynamics_rect}(c) and \ref{fig:dynamics_rect}(e), the mean-field Eqs.~\eqref{m_f_eq_Rect} with $n=10$ approximate well the dynamics of $N=5\times 10^4$ QIF noisy neurons with uniformly distributed heterogeneity. 

\section{\label{sec:Conclusions} Conclusions}

We investigated the influence of independent Cauchy white noise on the dynamics of large populations of inhibitory QIF neurons with synaptic kinetics and two types of quenched heterogeneities defined by a family of $q$-Gaussian distributions and a family of flat distributions. Both families depend on a positive integer parameter $n$, so that for $n=1$ they coincide with the Cauchy distribution, but for $n \to \infty$ the first family evolves towards a normal distribution, while the second tends to a uniform distribution. Using the Ott-Antonsen~\cite{Ott2008} ansatz, finite-dimensional systems of mean-field equations for both types of heterogeneities are obtained. These systems accurately describe the asymptotic dynamics of QIF neurons in the thermodynamic limit. We performed a detailed bifurcation analysis of the mean-field equations for both types of quenched heterogeneities. The regions of stable incoherent equilibrium states and coherent limit cycle oscillations were established in different planes of system parameters.

A similar system of mean-field equations was obtained earlier for the case of Cauchy noise and Cauchy quenched heterogeneity~\cite{Clusella2022}. It turned out that the effects of noise and heterogeneity are summed in these equations so that a change in noise intensity by some amount is equivalent to a change in quenched heterogeneity by the same amount. Here we generalized these equations to non-Cauchy heterogeneity and showed that noise and heterogeneity in this case have different effects on the collective dynamics of neurons. The existence of coherent limit cycle oscillations proved to be less sensitive to an increase in heterogeneity than to an increase in noise intensity. More interestingly, we found that changes in noise and heterogeneity in opposite directions can lead to qualitatively the same effect. In particular, collective limit cycle oscillations can be induced by either a reduction in noise or an increase in quenched heterogeneity. We have demonstrated this nontrivial effect for both types of quenched heterogeneities defined by a family of $q$-Gaussian distributions and a family of flat distributions. 

To verify the validity of the mean-field equation obtained in this paper, we compared their solutions with the solutions of the corresponding large-scale stochastic neural networks. The collective dynamics of a network consisting of $5\times 10^4$ QIF neurons with normally distributed quenched heterogeneity agrees well with the solution of the mean-field equations derived for a family of  $q$-Gaussian distributions with  MTI $n=10$. Also, the collective dynamics of the same size network with uniformly distributed quenched heterogeneity is well approximated by the mean-field equations obtained for a family of flat distributions with $n=10$.

In this paper, as in Ref.~\cite{Clusella2022}, we used the standard Ott-Antosen theory to derive mean-field equations for the case of Cauchy noise and non-Cauchy quenched heterogeneity. In a recent publication~\cite{Pietras2023}, the authors developed a new approach that goes beyond the OA theory. For the case of Cauchy noise and Cauchy quenched heterogeneity they derived a more general version of the mean-field equations, which, unlike the equations  in~\cite{Clusella2022}, can describe not only the dynamics on the OA manifold but also the transient dynamics for initial conditions outside this manifold. The authors in~\cite{Pietras2023} noted that their approach can also be generalized to non-Cauchy quenched heterogeneity.

\section*{Acknowledgments}

This work is supported by grant No. S-MIP-21-2 of the Research Council of Lithuania. The authors thank Alessandro Torcini for the helpful discussions.

\bibliographystyle{elsarticle-num}
\bibliography{QIF_Cauchy_Noise}
\end{document}